\documentclass[aps,prb,twocolumn,superscriptaddress,superscriptaddress,longbibliography]{revtex4-1}
\usepackage{epsfig,amsopn}
\usepackage{graphicx}
\usepackage{color}
\usepackage{amsmath,amssymb}
\usepackage{enumerate}
\newcommand\bea{\begin{eqnarray}}
\newcommand\eea{\end{eqnarray}}
\newcommand\beq{\begin{equation}}
\newcommand\eeq{\end{equation}}

\def\nn{\nonumber}

\def\si{\sigma}

\def\De{\Delta}
\def\dg{\dagger}

\def\ra{\rangle}
\def\ua{\uparrow}
\def\da{\downarrow}

\def\ka{\kappa}

\begin{document}
\title{Tunable crossed Andreev reflection in a heterostructure consisting of ferromagnets, normal metal and superconductors } 
\author{ Abhiram Soori~~}
\email{abhirams@uohyd.ac.in}
\affiliation{ School of Physics, University of Hyderabad, C. R. Rao Road, Gachibowli, Hyderabad-500046, India.}
\begin{abstract}
Crossed Andreev reflection (CAR) is a nonlocal transport phenomenon in a system of two normal metal (NM) leads connected to a superconductor (SC) that converts the electron like excitations in one metallic lead into hole like excitations in the other metallic lead. The scattering phenomena viz. electron tunneling (ET), electron reflection (ER) and Andreev reflection (AR) compete with CAR and reduce the probability of CAR generically. One of the experimentally realized proposals to observe CAR is to employ two ferromagnetic (FM) leads in antiparallel configuration connected to the SC by suppressing ET and AR. But CAR probability cannot be tuned in this setup. We propose a setup consisting of a gate tunable NM region connected to two superconducting regions on either side which are connected to FM leads  further away in antiparallel configuration, in which probabilities of CAR and ER can be changed from $0$ to $1$ by a changing the  gate voltage applied to the NM region. The gate voltage applied to the NM region dictates the chemical potential of the NM region and gives a handle on the Fabry-P\'erot interference of the electron and hole modes in the NM region. We calculate differential transconductance for the proposed setup which can be tuned across the range $0$ to  $-e^2/h$.
\end{abstract}
\maketitle
\section{Introduction}
Crossed Andreev reflection~(CAR) is a nonlocal transport phenomenon that happens in a system of two normal metal~(NM) leads connected to a superconductor~(SC) where an electron current driven in one NM lead results in a hole current exiting through the other NM lead~\cite{deutscher2000,cht,melin2004sign,bovzovic2002coherent,dong2003coherent,yamashita2003crossed,beckmann2004evidence,beckmann2006d,russo2005experimental,yeyati07,linder2009spin,reinthaler2013proposal,linder2014superconducting,crepin2013even,he14,wang2015quantized,sadovskyy2015,soori17,nehra19}. This phenomenon is useful for many reasons, one being the production of nonlocally entangled electrons~\cite{recher2001andreev,bena2002quantum,das2012high,schindele2012near}, another as a probe for nonlocality of Majorana fermions~\cite{fu10,hutzen12,wang13,sau15,law09,soori19,soori20ssc}.  CAR in a system of two NM leads connected to a SC is accompanied by other competing processes where the electron current driven in one NM lead results in :(i) an electron current exiting the same NM lead, (ii) a hole current exiting the same NM lead, or (iii) an electron current exiting the other NM lead; these phenomena are  called electron reflection (ER), Andreev reflection~(AR) and electron tunneling~(ET) respectively. Current due to ET in the second NM is typically much larger than the current due to CAR, in addition to the two currents being opposite in sign and this hinders the observation of CAR~\cite{cht,soori17}. Hence, enhancement of CAR over ET is essential for CAR to be observed. Of many proposals to enhance CAR over other competing processes~\cite{deutscher2000,cht,melin2004sign,sadovskyy2015,soori17}, two schemes have been realized experimentally so far. In one scheme, two ferromagnets in antiparallel configuration have been connected to the SC~\cite{beckmann2004evidence,beckmann2006d}, while in the other scheme, barriers have been employed at the NM-SC interfaces~\cite{russo2005experimental}. However, the strength of CAR current is not tunable in these experiments. A control over the magnitude of CAR enables us to enhance this scattering phenomenon by tuning the relevant parameter in the setup.  

Fabry-P\'erot interference on the other hand is a phenomenon that has played an important role in electron transport~\cite{liang2001}. This is essentially an interference between waves reflected multiple times in a ballistic channel between two partially transparent interfaces.  Fractional charges in quantum Hall systems have been detected using this phenomenon~\cite{ofek2010}. Oscillations in transverse conductance in the study of planar Hall effect are due to this phenomenon~\cite{suri21,soori2021phesoc}. Conductance oscillations in proposed spin transistors based on edge states of quantum spin Hall insulators have been interpreted with the help of  this phenomenon~\cite{soori12}. One of the schemes to enhance CAR is based on Fabry-P\'erot interference between Bogoliubov-de Genes quasiparticles formed in superconducting ladder~\cite{soori17,nehra19}. By applying bias across both the NM-SC junctions, relative contributions from CAR and ET can be tuned by changing the two bias voltages when the SC region also has spin orbit coupling~\cite{wu14}. 

\begin{figure*}
\begin{center}
  \includegraphics[width=12cm]{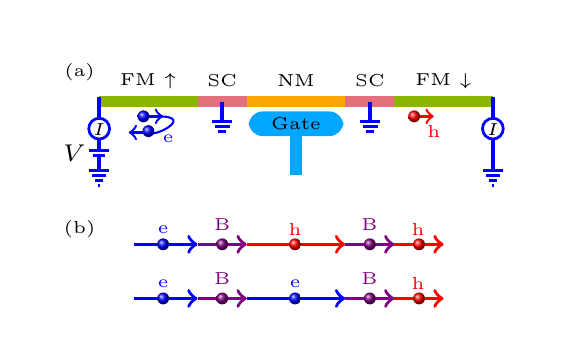}
  \end{center}
 \caption{(a) A schematic of the proposed setup. Two ferromagnetic leads~(FM) in antiparallel configuration placed at the two extreme ends are connected to the central gate tunable normal metal~(NM) region via superconductors~(SCs). Both the FMs are fully spin polarized. A bias voltage $V$ is applied between the left FM and the right FM, and the currents are measured in both the leads. An electron~(e) incident from the left FM either gets reflected as an e in the left FM or gets transmitted as a hole~(h) into the right FM.
 (b) An illustration of prominent scattering processes that interfere and result in crossed Andreev reflection. In both processes illustrated here, an incident electron from left FM exits as a hole in the right FM. In the central NM region, it traverses as an electron in one of the processes and as a hole in the other process as shown. In the SC regions, the excitation travels as a BdG quasiparticle~(B). }\label{fig:schem}
\end{figure*}

We propose a setup to enhance CAR by Fabry-P\'erot interference between electron and hole states of a gate tunable NM region connected to superconductors on either side which are connected to ferromagnetic leads further away in  antiparallel configuration as depicted in Fig.~\ref{fig:schem}. In our setup, the ferromagnetic leads in antiparallel configuration are fully spin polarized  and hence AR and ET are completely suppressed. The superconductor is an s-wave superconductor which mixes electron and hole excitations in opposite spin channels. Hence, an electron incident from left FM lead either gets reflected as an electron in left FM lead (ER) or gets transmitted as a hole in right FM lead (CAR). Whether CAR dominates ER or the other way is decided by the interference among the electron and hole excitations in the middle NM region which can be tuned by an applied gate voltage. 

This paper is organized as follows. In section~\ref{sec:calc},  we present the Hamiltonian for the proposed system and expatiate on the calculations. In section~\ref{sec:res}, we present and understand the results obtained. In section~\ref{sec:con}, we discuss  possible experimental realization of the proposed setup, future directions and conclude.

\section{Details of Calculation}\label{sec:calc}
We describe the system by the following Hamiltonian on a one dimensional lattice
\bea H &=& -t\sum_{n=-\infty}^{\infty}[\Psi^{\dg}_{n+1}\tau_z\Psi_n +{\rm h.c.}] \nn \\ &&-\mu\Big(\sum_{n=-\infty}^{-1}+\sum_{n=N_0+1}^{\infty}\Big)\Psi^{\dg}_n\tau_z\Psi_n -M\sum_{n=-\infty}^{-N_S-1}\Psi^{\dg}_n\si_z\Psi_n \nn \\ && +\De\Big(\sum_{n=-N_S}^{-1}+\sum_{n=N_0+1}^{N_0+N_S}\Big)\Psi^{\dg}_n\tau_x\Psi_n  -\mu_0\sum_{n=0}^{N_0}\Psi^{\dg}_n\tau_z\Psi_n \nn \\ &&+ M\sum_{n=N_0+N_S+1}^{\infty}\Psi^{\dg}_n\si_z\Psi_n,  \label{eq:H} \eea
where $t$ is the hopping strength in all the regions, $\mu$ is the chemical potential in the FM leads and the SCs, $M$ is the Zeeman energy that characterizes the FMs, $N_S$ is the number of sites for each of the SC regions, $\De$ is the strength of superconducting pairing, $\mu_0$ is the chemical potential of the central NM region, $(N_0+1)$ is the number of sites in the NM region, $\Psi_n=[c_{\ua,n}, c_{\da,n}, -c^{\dg}_{\da,n}, c^{\dg}_{\ua,n}]^T$, $c_{\si,n}$ and $c^{\dg}_{\si,n}$ are annihilation and creation operators respectively for electron with spin $\si$ at site $n$, $\si_z$ is the Pauli spin matrix acting on the spin space, and $\tau_x, \tau_z$ are the Pauli spin matrices acting  on the particle-hole space. On the left FM (in the region $n\le -N_S-1$), the dispersion relations for up/down-spin electrons is $E=(-2t\cos{ka}-\mu \mp M)$ while the dispersion for up/down spin holes is $E=(2t\cos{ka}+\mu \mp M)$, where $a$ is the spacing between neighboring lattice points. If $(-V_0,V_0)$ is the bias window in which transport is studied, for $-2t<\mu<0$ and $V_0+2t+\mu<M<2t-\mu-V_0$, only up spin bands for electron and hole fall in the range of bias window in the left FM. The dispersions for right FM are the same except for the change $M\to-M$ due to antiparallel configuration. In the bias window, there are only down spin bands for electrons and holes. The dispersion in the two SC regions is $E=\pm\sqrt{(2t\cos{ka}+\mu)^2+\De^2}$. The dispersion in the central NM region is $E=\mp(2t\cos{ka}+\mu_0)$ for electron/hole bands of both the spins. The fact that chemical potentials of the superconductors do not change with time implies that the superconductors are grounded. 

An electron incident on the superconductor can get scattered either as an electron of same spin or a hole of opposite spin. There is no term in the Hamiltonian that mixes the spins. So, the wavefunction for an electron incident from the left FM lead has only two components that are nonzero: the up spin electron and the down spin hole. Such a wavefunction at energy $E$ has the form 
\bea \psi_n &=& (e^{ik_ean} + r_e e^{-ik_ean})|\ua_e\ra + r_h e^{\ka_han}|\da_h\ra,\nn \\&& ~~{\rm for~}n\le-N_S-1, \nn \\ &=& \sum_{j=1}^4s_{L,j}e^{ik_jan} [u, v_j]^T, ~~{\rm for~}-N_S\le n\le -1,\nn \\ &=& \sum_{s=\pm} [s_{0es}e^{sik_{e0}an}|\ua_e\ra + s_{0hs}e^{-sik_{h0}an}|\da_h\ra], \nn \\ &&{~~\rm for~}0\le n\le N_0, \nn \\ &=& \sum_{j=1}^4s_{R,j}e^{ik_jan} [u, v_j]^T, {~~\rm for ~} N_0+1\le n \le N_0+N_S, \nn \\ &=&t_e e^{-\ka_ean} |\ua_e\ra + t_h e^{-ik_han}|\da_h\ra, \nn \\ && ~~{\rm for~} n\ge N_0+N_S+1, \label{eq:psi}  \eea
where $k_ea=\cos^{-1}[-(\mu+E+M)/2t]$, $\ka_ha=\cosh^{-1}[(E-\mu+M)/2t]$, $|\ua_e\ra=[1,0]^T$, $|\da_h\ra=[0,1]^T$, $k_{1/2}a=\pm\cos^{-1}[-(\mu+\sqrt{E^2-\De^2})/2t]$, $k_{3/4}a=\pm\cos^{-1}[-(\mu-\sqrt{E^2-\De^2})/2t]$, $u=\De$, $v_j=(E+\mu+2t\cos{k_ja})$, $k_{e0}a=\cos^{-1}[-(\mu_0+E)/2t]$, $k_{h0}a=\cos^{-1}{[-(\mu_0-E)/2t]}$, $\ka_ea=\cosh^{-1}[-(\mu+E-M)/2t]$ and $k_ha=\cos^{-1}[-(\mu-E+M)/2t]$. While the eigenmodes in the FM and NM regions are either purely electron-like or purely hole-like, the eigenmodes in the SC region are the BdG quasiparticle modes which have both electron and hole components dictated by the spinor $[u,~v_j]^T$. When $|E|<\Delta$, the BdG quasiparticle modes are equally electron-like and hole-like since $|u|^2=|v_j|^2$. When $|E|>\Delta$, $j=1,2$ correspond to electron-like excitations while $j=3,4$ correspond to hole-like excitations. Further, $k_1=-k_2$ and $k_3=-k_4$ which imply that $k_{j}$'s come in pairs with one left-moving mode $j$ and one right-moving mode $j'$ such that $k_{j'}=-k_j$. The scattering coefficients $r_e$, $r_h$, $s_{L,j}$'s, $s_{0es}$'s, $s_{0hs}$'s, $s_{R,j}$'s, $t_e$ and $t_h$ have to be determined from Schr\"odinger wave equation for the Hamiltonian eq.~\eqref{eq:H}. Typically, for scattering problems in a continuum, boundary conditions need to be specified at the interfaces~\cite{soori17,soori2020,carreau90}. However, the system has been described by a lattice model in this work and the boundary conditions are encoded within the Hamiltonian eq.~\eqref{eq:H}. The Hamiltonian used in this work corresponds to highly transparent barriers at the interfaces~\cite{soori2020}. The reflected hole and the transmitted electron  modes are evanescent waves in the two FMs since there are no bands for these modes in the bias window. Two prominent scattering processes that contribute to CAR are illustrated in Fig.~\ref{fig:schem}(b). The amplitudes for ER and CAR- $r_e$ and $t_h$ must satisfy: $|r_e|^2+|t_h|^2\sin{k_ha}/\sin{k_ea}=1$, which follows from conservation of probability current. The differential transconductance $G_{RL}$ which is the ratio of infinitesimal change of current in the right FM lead to the infinitesimal change in voltage bias applied to the left FM lead at voltage bias $V=E/e$ is given by $G_{RL}=-|t_h|^2(e^2/h)\sin{k_ha}/\sin{k_ea}$, where $e$ is electron charge. Since the two SC regions are grounded, currents flow into the drains at the two SC's and are carried by Cooper pairs into the SCs. The steady state currents in the two FM leads and the NM region are not equal when a bias is applied. From the probability conservation, it can be shown that the local conductance at the left FM lead $G_{LL}$ (defined as the ratio of infinitesimal change of current in the left FM lead to the infinitesimal change in voltage bias applied to the left FM lead at voltage bias $V=E/e$) and the transconductance $G_{RL}$ are related by $G_{RL}=-G_{LL}$.

\section{Results and analysis}~\label{sec:res}
\begin{figure}
 \includegraphics[width=8cm]{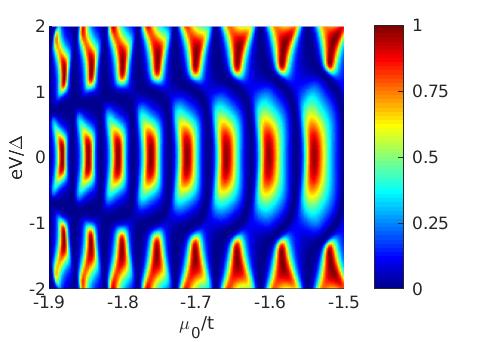}
 \caption{$|G_{RL}|$ the magnitude of differential transconductance plotted in units of $e^2/h$ as a function of chemical potential of the NM region $\mu_0$ and the bias for  $N_S=10$. Other parameters: $\mu=-1.9t$, $M=0.3t$, $\De=0.02t$ and $N_0=60$. }~\label{fig:GRLcontour}
\end{figure}

We set $\mu=-1.9t$ so that the Fermi energy is near the band bottom. The superconducting pairing $\De$ is typically much smaller than the difference between Fermi energy and the band bottom which is $0.1t$. So, we choose $\De=0.02t$. The Zeeman energy characterizing the ferromagnetic lead is chosen  to be $M=0.3t$ so that band of only one spin exists in the range of bias window $(-2\De,2\De)$. $(N_0+1)=61$ sites are chosen in the central $NM$ region. The transconductance is negative since only holes get transmitted into the right FM in response to electrons incident from the left FM. We plot a contour plot of magnitude of transconductance as a function of chemical potential of the central region $\mu_0$ and the voltage bias $V$ in Fig.~\ref{fig:GRLcontour} for an optimal choice of number of superconducting sites $N_S=10$. 
We find that $|G_{RL}|$ oscillates as $\mu_0$ is varied and at certain values of $\mu_0$ and bias $V$, it attains the maximum  value of $e^2/h$. In Fig.~\ref{fig:GRLcontour}, at zero bias, we find a series of values of $\mu_{0,i}$ at which $|G_{RL}|$ peaks. This is a feature of Fabry-P\'erot interference. To quantitatively uphold this explanation, at each value of $\mu_{0,i},~~i=1,2,..$, we find the wavenumbers for electron and hole excitations in the central NM region: $k_{e0,i}=k_{h0,i}=\cos^{-1}[-\mu_{0,i}/2t]$. These wavenumbers  satisfy the Fabry-P\'erot interference condition $(k_{e0,i+1}-k_{e0,i})(N_0+1)a=(k_{h0,i+1}-k_{h0,i})(N_0+1)a=\pi$ confirming that the oscillation in the transconductance is due to interference among electron and hole modes in the NM region. 

\begin{figure}
 \includegraphics[width=4.2cm]{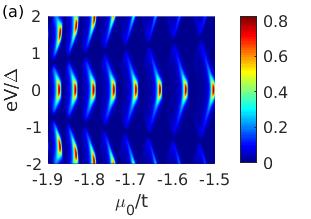}
 \includegraphics[width=4.2cm]{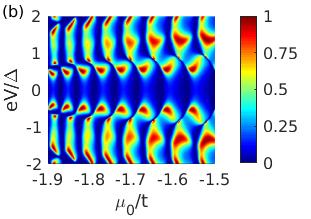}
 \caption{Magnitude of transconductance versus $\mu_0$ and bias for (a) $N_S=4$ and (b) $N_S=30$. Other parameters are same as in Fig.~\ref{fig:GRLcontour}. }\label{fig:GRL-><}
\end{figure}

We now turn to the conductance spectrum for lengths $N_S$ of the superconducting regions much smaller than and much larger than the optimal length. In Fig.~\ref{fig:GRL-><}, we plot the magnitude of transconductance versus $\mu_0$ and bias $V$ for (a)~$N_S=4$ and (b)~$N_S=30$. For very short SC regions, the CAR probability is small~\cite{cht,soori17} and hence the regions with enhanced CAR are narrow. Further, in the SC Hamiltonian, the magnitude of the term $\Delta$ that mixes the electron-like and hole-like particles is relatively larger than the magnitude of the term $(2t\cos{ka}+\mu)$ that corresponds to the pure NM close to zero bias. While the term in the Hamiltonian proportional to $\Delta$ converts an electron into hole, the term $t$ makes an electron at one site hop on to the neighboring site as an electron.  Hence, CAR probability is enhanced near zero bias for small values of $N_S$. However, there is a competing mechanism which takes over at large lengths of the SC region. Within the SC gap the wavenumbers $k_j$ in the SC regions are complex and the corresponding decay lengths are larger close to zero bias making transmission through the SC region harder close to zero bias. This can be seen from earlier works on transport across SC where both ET and CAR are suppressed close to zero bias for large lengths of the SC region~\cite{cht,soori17}. Suppression of CAR close to zero bias when $N_S$ is large is  rooted in this mechanism. This explains the contrasting behaviors of the transconductance at large and small lengths of the SC. 
The optimal length $N_{S0}\simeq[|{\rm Im}[\pi/(k_{\De}a)|]$, where $k_{\De}a=\cos^{-1}\{-(\mu+i\De)/2t\}$ is the wavenumber of one of the SC modes at zero energy and $[x]$ is the integer part of the real number $x$. For the values of parameters chosen for Fig.~\ref{fig:GRLcontour}, the optimal length of the SC: $N_{S0}\sim10$. 

\begin{figure}
 \includegraphics[width=4.2cm]{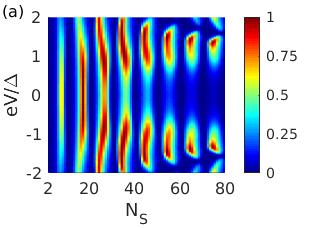}
 \includegraphics[width=4.2cm]{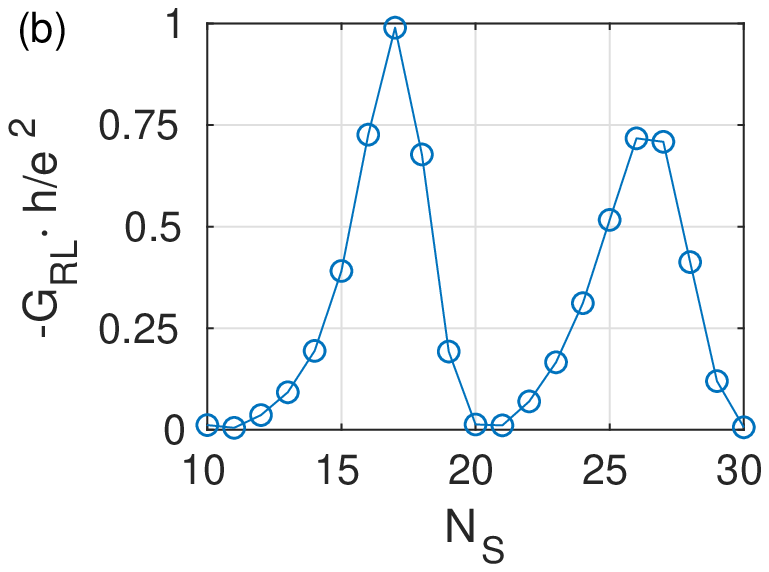}
 \caption{Dependence of the transconductance of FM-SC-FM heterostructure on length of the SC region $N_S$. (a) $|G_{RL}|$ in units of $e^2/h$ versus $N_S$ and $eV/\Delta$, (b) $|G_{RL}|$ at zero bias in units of $e^2/h$ versus $N_S$. Parameters chosen are same as in Fig.~\ref{fig:GRLcontour}. }\label{fig:FSF}
\end{figure}

To get an idea of enhancement of CAR in this scheme compared to FM-SC-FM heterostructures, we calculate $G_{RL}$ for an FM-SC-FM system with the SC region having $N_S$ sites. The Hamiltonian is the same as in eq.~\eqref{eq:H}, except that the central $NM$ region is absent and the SC region has totally $N_S$ sites. The magnitude of differential transconductance $|G_{RL}|$ is plotted  as a function of $N_S$ and bias in Fig.~\ref{fig:FSF}~(a) for the same choice of parameters as earlier. The magnitude of zero bias transconductance in the range of $10\le N_S\le 30$, where it is peaked is plotted in Fig.~\ref{fig:FSF}~(b). It can be seen from this figure that CAR is maximally enhanced for some special choice of length of the SC region: $N_S=17$. However the value of transconductance is highly sensitive to the choice of $N_S$. In an experimental system, the length of the SC region may not be precisely the one for which CAR is maximally enhanced and once the heterostructure is fabricated, the length of the SC region cannot be changed. 

\begin{figure}
 \includegraphics[width=7cm]{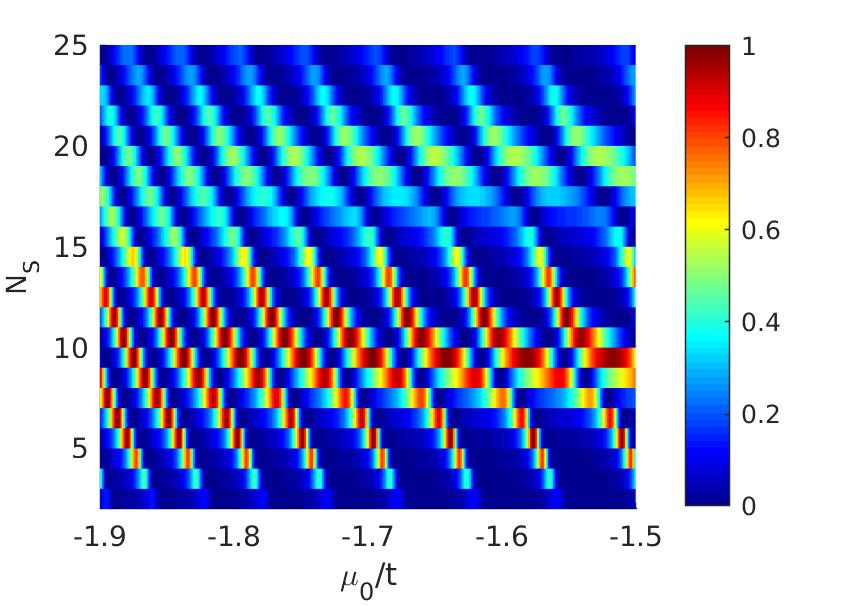}
 \caption{$|G_{RL}|$ at zero bias in units of $e^2/h$ versus $\mu_0$ and $N_S$ for the proposed FM-SC-NM-SC-FM setup. Parameters chosen are same as in Fig.~\ref{fig:GRLcontour}. }\label{fig:G0NS}
\end{figure}

To compare the sensitivity of $G_{RL}$ for the proposed FM-SC-NM-SC-FM system, we plot the magnitude of transconductance for this system at zero bias versus $\mu_0$ and $N_S$ in Fig.~\ref{fig:G0NS}. 
It can be seen from this figure that the value of $G_{RL}$
 is tunable to maximal value close to $e^2/h$ by changing $\mu_0$ for a range of lengths of SC: $5<N_S<13$.

\section{Discussion and conclusion}~\label{sec:con}
We have proposed a scheme where CAR can be tuned and enhanced. The proposed scheme encompasses a gate tunable normal metal region connected to FM leads through SC regions. Spin orbit coupling in the system induces spin flip scatterings and the incident up spin electron can get transmitted as a down spin electron in the proposed setup adding another scattering phenomenon that competes with CAR. Hence, an ideal candidate for the NM region should have negligible spin orbit interaction. Experimentally, chemical potential can be tuned by an applied gate voltage in graphene~\cite{castroneto09,liao10} -a material with negligible spin orbit interaction making it an ideal candidate for NM region.   A few candidates for fully spin polarized ferromagnets such as  perovskites~\cite{Park1998} and CrO$_2$~\cite{singh2015} exist, in addition to ferromagnets like iron in which a substantial spin polarization can be induced by an applied magnetic field. AR has been observed in junctions of graphene with NbSe$_2$~\cite{efetov2016,sahu16}, making NbSe$_2$ a possible choice for SC in the proposed setup. 

CAR is enhanced maximally in certain regions of the parameter space. Such regions of maximally enhanced CAR are substantially large for a range of length of the SC region which is determined by the SC pairing strength, the chemical potential of the SC region and the hopping strength. The probabilities for CAR and ER can be tuned by an  applied gate voltage which changes the chemical potential of the NM region in the setup. The Fabry-P\'erot interference condition: $(k_{e0,i+1}-k_{e0,i})(N_0+1)=\pi$ implies the spacing between consecutive values of $\mu_0$ at which CAR probability is maximum gets smaller as the number of NM sites $N_0$ increases. For small values of $N_0$, $\mu_0$ has to be changed substantially to alter the CAR probability by a considerable amount whereas for large $N_0$ even a tiny change in $\mu_0$ affects the interference resulting in ample variation of CAR probability. Hence, a choice of $N_0$ neither too small nor too large makes the tunability of CAR practically possible. 

Effects such as Coulomb blockade and inelastic scattering can be important in heterostructures considered~\cite{giazotto}. Coulomb blockade is a phenomenon that originates from electron electron interaction in the NM region of such a setup. By choosing a large capacitance for the gate electrode coupled to the NM region, the effects of Coulomb blockade can be suppressed~\cite{ihn}. Inelastic scattering at NM-SC junctions has explained Andreev reflection features in certain systems with the help of a formalism which uses Dynes parameter~\cite{dynes84,plecen94,chen12}. However, this formalism cannot be extended to nonlocal transport across a superconductor to get physically meaningful results. Further, the phenomenon of CAR observed in a number of experiments~\cite{beckmann2004evidence,beckmann2006d,russo2005experimental,devries19} has been satisfactorily explained using the scattering formalism employed in this work without taking into account inelastic scattering. Developing a formalism to study nonlocal transport across a superconductor taking inelastic relaxation into consideration is a promising future direction. We believe our proposal can be realized with the present technology by suitably fabricated heterostructures. 

\acknowledgements
The author thanks Subroto Mukerjee and Dhavala Suri for useful discussions, and DST-INSPIRE Faculty Award (Faculty Reg. No.~:~IFA17-PH190) for financial support. 
\bibliography{ref_car}
\end{document}